\documentclass[aps,prc,superscriptaddress,nofootinbib,showpacs,floatfix,reprint]{revtex4-1}
\usepackage[utf8]{inputenc}
\usepackage{graphicx} % Required for inserting images
\usepackage{amsmath} 
\usepackage{amsfonts}
\usepackage{amssymb}
\usepackage{amsthm}
\usepackage{bm}
\usepackage{xcolor}
\usepackage[english]{babel}
\usepackage{fontenc}
\usepackage[margin=1in]{geometry}
\usepackage{enumerate}
\usepackage{color}
\usepackage{lineno}
\definecolor{darkgreen}{rgb}{0.0, 0.6, 0.22}
\definecolor{dblue}{rgb}{0.,0.,1.}       
\definecolor{orange}{cmyk}{0.,0.453,1.,0.}    % orange
\begin{document}
\title{Correlations of Feed-down Hadrons in a Thermal Model}

\author{ Claude~Pruneau }
\email{claude.pruneau@wayne.edu}
\author{ Victor~Gonzalez }
\email{victor.gonzalez@cern.ch}
\author{ Oveis~Sheibani }
\email{oveis.sheibani@cern.ch}
\author{ Chun~Shen}
\email{chunshen@wayne.edu}
\affiliation{Department of Physics and Astronomy, Wayne State University, Detroit, 48201, USA}
\author{Yash Patley}
\email{yashpatley.iitb@gmail.com}
\author{Basanta Nandi }
\email{basanta@phy.iitb.ac.in}
%\author{ Sadhana Dash }
%\email{sadhana@phy.iitb.ac.in}
\affiliation{Department of Physics, Indian Institute of Technology Bombay, Mumbai - 400076, India }
\author{Ana Marin} 
\email{a.marin@gsi.de}
\affiliation{GSI Helmholtzzentrum f\"ur Schwerionenforschung GmbH, Research Division and ExtreMe Matter Institute EMMI, 64291 Darmstadt, Germany}

\date{\today}

\begin{abstract}
We examined the potential impact of decays on the magnitude of fluctuations of net quantum numbers and integrals of balance functions based on a thermal hadron gas model. The calculations are based on a comprehensive list of known hadrons with masses up to 2.5 GeV/$c^2$ and include all decays of these hadrons with known branching fractions. The calculations are performed at vanishing baryo-chemical potential for temperatures between 140 and 200 MeV. We show that the decays feed-down substantially impact the single yield of measurable  ``stable particles" as well as those of correlated densities of these species. Decays can then potentially have large and non-trivial impacts on measurements of net quantum number cumulants and balance functions. These observations are particularly important in the context of the search for the QCD critical point at the  RHIC Beam Energy Scan as well as efforts to determine chemical susceptibilities near the phase transition at RHIC or LHC energies. Results obtained in this work also shed light on the importance of feed-down in measurements of balance functions in elementary p-p and nucleus-nucleus collisions.
\end{abstract}
\maketitle

\section{Introduction}
\label{sec:introduction}

%Andronic:2020iyg,Braun-Munzinger:2020jbk,Braun-Munzinger:2016yjz,

Measurements of fluctuations of the net conserved quantum numbers of produced particles are of particular interest because they are related to fundamental properties of the nuclear matter produced in nucleus-nucleus (A-A) collisions at high energy~\cite{An:2022jgc,Parra:2025fse,Borsanyi:2025ttb}. Cumulants of net conserved quantum numbers, in particular, are directly proportional to the nuclear matter susceptibilities near the phase transition~\cite{Braun-Munzinger:2020jbk,Andronic:2020iyg,Borsanyi:2025ygf,Friman:2011pf}. They have also been exploited to seek evidence for a critical point (CP) of nuclear matter in the context of the beam energy scan (BES) at the Relativistic Heavy Ion Collider (RHIC)~\cite{Stephanov:2009ra,An:2021wof,Borsanyi:2025dyp,STAR:2023zhl,STAR:2022etb,STAR:2022vlo,STAR:2021fge,STAR:2021iop,Pratt:2020ekp}.

Similarly, measurements of correlations functions, specifically charge and baryon number balance functions (BFs), are of interest because they are highly sensitive to  the production and transport of particles in the midst of A--A collisions~\cite{Pratt:2022kvz,Pratt:2023pee} and properties of the created medium, such as the light quark diffusivity~\cite{Pratt:2019pnd,Pratt:2021xvg,Gonzalez:2020bqm}. Charge balance functions, in particular, were in fact proposed as a tool to probe evidence of delayed hadronization in A--A collisions and the formation of quark gluon plasma (QGP) matter~\cite{Pratt:1999ku}. 

However,  cumulants of conserved quantum numbers  and balance functions are both impacted by background processes likely to modify or even mask the phenomena sought for in these measurements. Background processes include global constraints associated with conservation laws~\cite{Pruneau:2019baa,Braun-Munzinger:2023gsd}, large feed-down (FD) from resonance decays, as well as jet production. The latter two are likely most impactful at top RHIC energy and LHC energy, whereas the effect of quantum number conservations are present at all beam energies. While many measurements of jet production cross sections and their properties have been performed both at RHIC and the LHC~\cite{STAR:2013thw,STAR:2017ieb,ALICE:2023oww}, relatively little attention has so far been given to the role of decays in these measurements. It is thus of interest to obtain baseline results on the impact of resonance decays of high-mass hadrons in measurements of integral (fluctuation cumulants) and differential correlations (balance functions).    

Resonance decays are important to consider because they explicitly introduce correlations between observed particles. 
Correlations between hadrons produced by FD stem from their common source, the parent particle, as well as energy-momentum conservation, which determines the momenta of the daughter particles in relation to their parent, and quantum number conservation. Resonance decays leave the net baryon number unchanged, yet they can strongly affect measurements that rely on observed protons and antiprotons as proxies for the total baryon number in fluctuation and correlation analyses. They also substantially modify the yields of light mesons such as pions and kaons. 

Nominally, the presence of FD of hadrons into pions or kaons plus other light hadrons can be identified and quantified based on the reconstruction of the invariant mass of high mass hadrons that decay into measured light hadrons. This technique has severe limits, however, because the multiplicity of particles produced in A--A collisions at RHIC and LHC is very large and thus generates combinatorial backgrounds that make the determination of parent yields extremely challenging in the context of both integral and differential correlation measurements. Additionally, although resonance decays can nominally be identified in measurements of balance functions expressed as a function of the invariant momentum $Q^2$ of daughter particles~\cite{Pratt:2022xbk}, this technique is limited to relatively narrow and easily identifiable resonances, e.g., a $\phi$-meson decaying into a pair of kaons ($\phi\rightarrow \rm K^+ + K^-$). In practice, most hadron resonances are broad and so closely packed together that their identification is impractical in fluctuations or correlations  measurements. An experimental determination of the impact of resonance decays thus seems rather challenging if not impossible at the outset. 

It might, however, be possible to seek guidance from theoretical considerations. For instance, in the context of elementary collisions, e.g., pp collisions, Monte Carlo models such as \textsc{Pythia} and HERWIG have been used with qualitative  success to predict the strength and shape of correlations~\cite{Sahoo:2018uhb,Pruneau:2024jpa}. By design, \textsc{Pythia} and HERWIG explicitly account for all conservation laws and they decay hadrons based on their measured properties.  Examining the importance of decay correlations in A--A collisions may be somewhat more challenging, given that these feature strong collective modes and description in terms of hydrodynamics do not, in general, account for quantum number conservation at the local fluid cell level.  One can nonetheless get some assessment of the roles of decays in fluctuations/correlations based on an idealized thermalized hadron gas.

The goal of this work is to estimate the impact of decay FD onto measurable ``stable particles" based on a simple version of the thermal hadron gas model~\cite{Andronic:2008ev}. Our discussion is framed in the context of the grand canonical model (GCM) and capitalizes on existing tools developed towards the estimation of the role of FD on single particle production~\cite{Chojnacki:2011hb}. It is assumed that hot QCD matter formed in A--A collisions cools down sufficiently slowly, as the collision system expands, to remain in thermal equilibrium at all times until chemical and kinetic freeze out. It is additionally assumed that all hadrons, irrespective of their mass and flavor content, reach chemical freeze-out at the same proper time, so one can neglect the fact that heavier particles stop being formed earlier (by detailed balance) than their lighter cousins. Effects of quantum number conservation have been addressed  by other authors and are not considered in this work~\cite{Vovchenko:2020gne,Braun-Munzinger:2023gsd}. 

\section{Method}

In the Grand Canonical Model (GCM), the (thermal) density $\rho_1^{\rm TH}(\alpha)$ of a species $\alpha$ of mass $m_{\alpha}$, maintained in a heat bath at temperature $T$ and chemical potential $\mu_{\alpha}$, is given by 
%\begin{widetext}
\begin{align}
\label{eq:1}
\rho_1^{\rm TH}(\alpha) &= \frac{g_{\alpha}}{2\pi^2}\int_0^{\infty}\frac{p^2{\rm d}p}{\exp\left[ \left(E_{\alpha}-\mu_{\alpha} \right)/T \right]\pm 1},
\end{align}
%\end{widetext}
where  $p$ and $E_{\alpha}$ are the momentum and energy of particles of species $\alpha$, and $g_{\alpha}$ is their spin degeneracy factor. Integration of Eq.~(\ref{eq:1}) yields
\begin{align}
\label{eq:2}%
        \rho_1^{\rm TH}(\alpha)=&\!\frac{g_{\alpha} T m_{\alpha}^2}{2\pi^2} \sum_{k=1}^{\infty} \frac{(-s)^{k+1}}{k} K_2\left( \frac{k m}{T}\right) \\ \nonumber 
        & \times \exp\left(\pm k \mu_{\alpha}/T\right),
\end{align}
in which $K_2(z)$ represents second order modified Bessel function, with $s=1$ for fermions and $s=-1$ for bosons. Although the sum  nominally proceeds from $k=1$ to infinity, one finds that it converges rapidly. A maximum value of $k=6$ was used in the calculations reported in this work. 

Our calculation of stable species densities is based on the particle and decay tables included in the distribution of the Therminator2~\cite{Chojnacki:2011hb} package. The species table  includes all known non-strange and strange hadrons up to mass 2.5 GeV/$c^2$ and the decay table includes their two, three, and four prong decays channels. Known branching fractions are used whenever possible and corrections for degrees of freedom are carried out with Clebsch-Gordon coefficients~\cite{Chojnacki:2011hb}. 

\begin{figure*}
\centering
\includegraphics[scale=0.25,trim={4mm 15mm 5mm 12mm},clip]{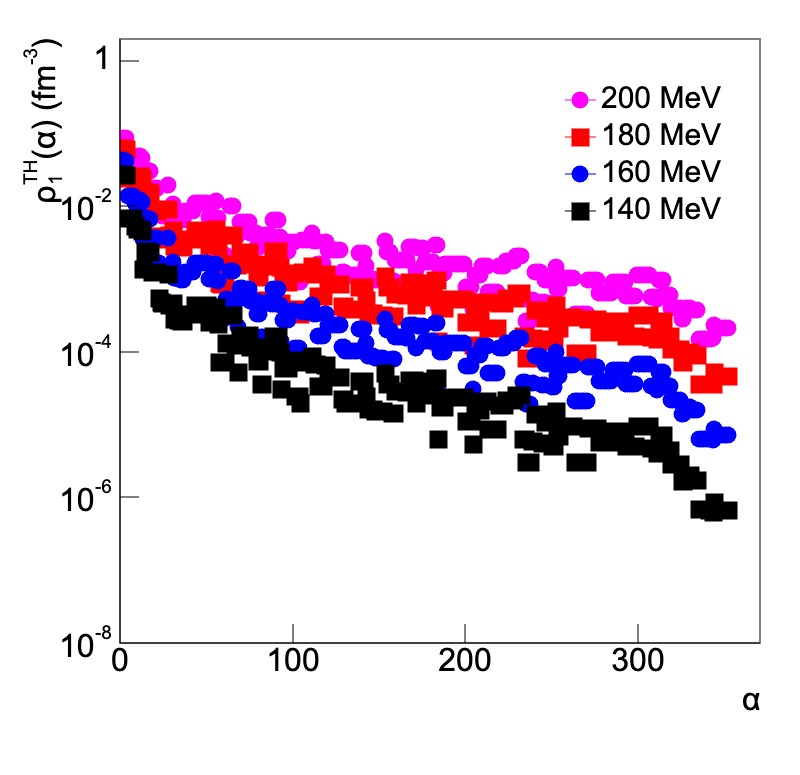}
\includegraphics[scale=0.25,trim={4mm 15mm 5mm 12mm},clip]{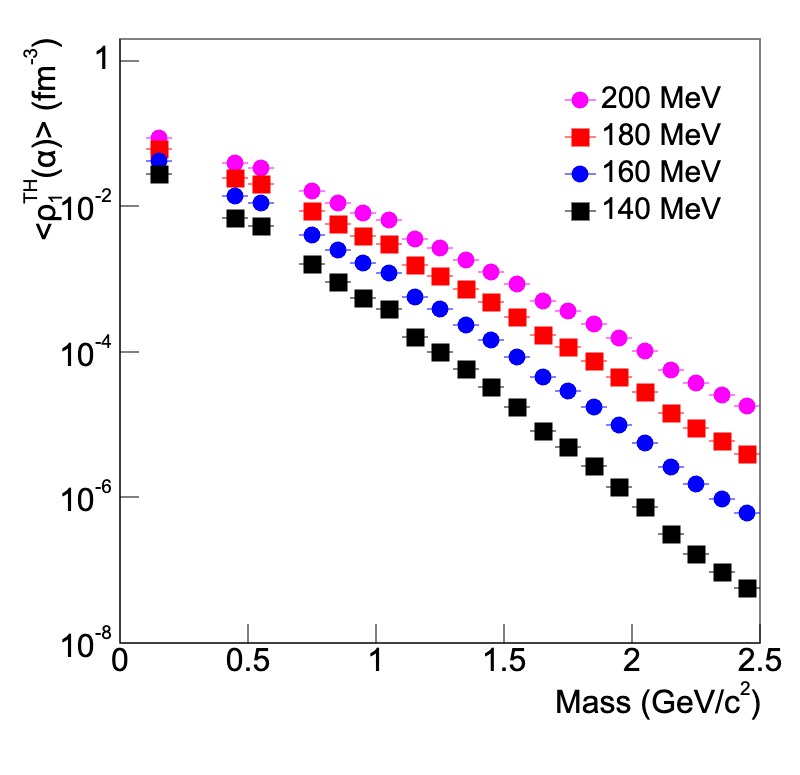}
\caption{ (left) Thermal density of individual hadron species plotted as a function of their index $\alpha\in[1,353]$; (right) Average thermal density vs. species mass. Densities are computed with the  thermal model presented in the text for selected temperatures and vanishing baryo-chemical potential.}
\label{fig:ThermalDensities}
\end{figure*}

In this work, particles designated as stable include pions ($\pi^{\pm,0}$), kaons ($\rm K^{\pm,0}, \overline{K}^0$), as well as protons ($\rm p$), neutrons ($\rm n$), lambda-baryon ($\Lambda^{0}$) and their respective anti-particles $\rm \bar{p}$, $\rm \bar{n}$, and $\overline{\Lambda}$. 
%, enhanced to include charm and bottom hadrons.   
All species from the particle table  are sorted in increasing order of mass and attributed a sequential arbitrary index $\alpha =1, \ldots, 353$. Charm and bottom hadrons are neglected given their expected low thermal yield and unresolved questions regarding their degree of thermalization. The left panel of Fig.~\ref{fig:ThermalDensities} displays a graph of the thermal densities $\rho_1^{\rm TH}(\alpha)$, for $\alpha=1,\ldots, 353$, computed for particles with mass smaller than 2.5 GeV/$c^2$ while the right panel shows the average densities plotted as a function of the mass of particles. Deviations from a smooth monotonic trend arise from the spin degeneracy factor $g_{\alpha}$. Densities are computed for vanishing chemical potentials for four representative temperatures $T=140$, 160, 180, and 200 MeV. One observes that the density of low-mass particles, i.e., pions, increases by as little as a factor of $\sim 3$, whereas the densities of the heaviest species considered rise by more than 200 times when changing the temperature from $T=140$ to $T=200$~MeV. 

The calculation of species densities resulting from FD proceeds in two steps based on the sorted mass of the known species. Given stable species $a$ do not decay, they are given a unit probability of decaying into themselves, $P^{a}_{a} = 1$. One then computes the probability of species $\alpha$ to decay into stable species $a$ iteratively starting with the lowest mass objects. For the calculation of single particle FD, one keeps track of all alternative paths/species, $i$ which produces the decay product $a$. The probabilities for the level $\alpha$ are then computed, for $i=1,\ldots, n'$, according to  
\begin{align} 
\label{eq:singleProb}
    P^{\alpha}_{a} &= \sum_{i=1}^{n'} B_{i}^{\alpha} P_{a}^{i},
\end{align}
with $\alpha=1,\ldots, n$, where $n$ is the total number of species and $n'$ is the total number of alternative paths/species which from species $\alpha$ produces $a$. $B_{i}^{\alpha}$ is the branching fraction of species $\alpha$ into a specific decay channel involving a species of type $i$, and $P^{i}_{a}$ is the probability of species $i$ decaying into a stable species $a$. Obviously, this step requires the different levels $i$ being available from previous iterations and enables the computation of the level $\alpha + 1$. Once the iterative process is concluded, the FD density of species $a$, denoted $\rho_{a}^{\rm FD}$, is computed according to 
\begin{align} 
\label{eq:singleFDDensity}
    \rho^{\rm FD}_{a} &= \sum_{\alpha=1}^{n} \rho^{\rm TH}_{\alpha} P_{a}^{\alpha},  
\end{align}
where $n=353$ is the number of species included in the calculation.
The total density of stable species $a$ is the sum of their thermal and FD densities
\begin{align} 
\label{eq:singleTHFDDensity}
    \rho^{\rm TH+FD}_{a} &= \rho^{\rm TH}_{a} + \rho^{\rm FD}_{a}.  
\end{align}

\section{Results}

\begin{figure}[hb]
\centering
\includegraphics[width=0.99\linewidth,trim={7mm 2mm 5mm 2mm},clip]{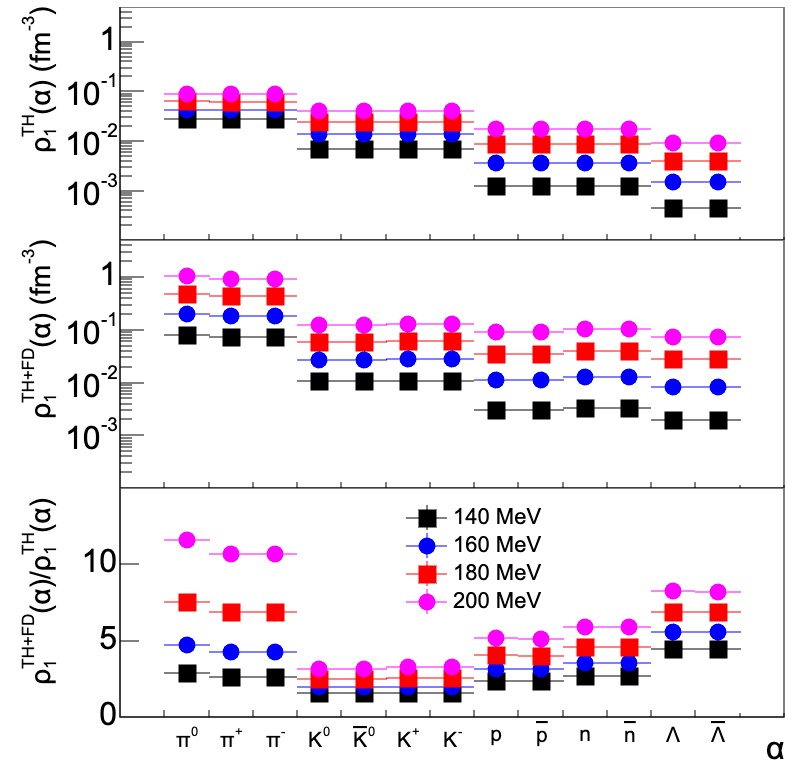}
\caption{(top) Thermal densities of stable hadrons computed at selected temperatures;(middle) densities obtained after adding FD from unstable higher mass hadrons; (bottom) Ratio of densities with and without FD.}
\label{fig:StableHadronSingleDensities}
\end{figure}

Figure~\ref{fig:StableHadronSingleDensities} presents the thermal, in the top panel, and thermal plus FD densities of stable particles computed with Eqs.~(\ref{eq:2},\ref{eq:singleFDDensity},\ref{eq:singleTHFDDensity}) in the middle panel. The bottom panel displays the ratio $\rho^{\rm TH+FD}_{\alpha}/\rho^{\rm TH}_{\alpha}$. Within this calculation, particles and their respective antiparticles have equal thermal yields because they have the same mass and chemical potentials are set to zero. Pions being the lightest hadrons, their total yields are found to increase the most with FD contributions, whereas kaon yields are the least impacted by FD. The relative importance of FD contributions increases  significantly with the temperature, rising for $\pi^+$ from a factor of 2.64 at $T=140$ MeV to 10.6 at $T=200$ MeV. 
Among baryons, protons are the least impacted by FD while the yield of $\Lambda$ increases the most at any given temperature. 

Focusing on the impact of FD at $T=$ 160 MeV, which corresponds to typical freeze-out temperatures obtained by blast wave model fits of single species momentum spectra at both RHIC (top energy) and LHC~\cite{Xu:2017akx}, one notes that FD contributions increase proton density by a factor of $\sim 3.1$. This is relevant to note given that measurement of net-baryon yield fluctuations, particularly net-protons, are used as a proxy for net-baryon fluctuations aimed at a determination of chemical susceptibilities and searches for the critical point of nuclear matter.  Although, strong and weak decays preserve the net baryon number, they do not preserve the net proton number: baryon decays feature p, n, or $\Lambda$ as end points (in the context of this discussion). 
The density of observable  protons $\rho_{\rm p}^{\rm TH+FD}$, assuming perfect efficiency and acceptance, is thus related to that of the total primary baryon density $\rho_{\rm B}^{\rm TH}$ according to

\begin{align}
\rho_{\rm B}^{\rm TH} &= \left( 1 + \frac{\rho_{\rm n}^{\rm TH+FD}}{\rho_{\rm p}^{\rm TH+FD}}
+ \frac{\rho_{\Lambda}^{\rm TH+FD}}{\rho_{\rm p}^{\rm TH+FD} } \right) \rho_{\rm p}^{\rm TH+FD}  \\ \nonumber
&= 2.87 \, \rho_{\rm p}^{\rm TH+FD},
\end{align}
where the ratios $\rho_{\rm n}^{\rm TH+FD}/\rho_{\rm p}^{\rm TH+FD}=1.13$ and $\rho_{\Lambda}^{\rm TH+FD}/\rho_{\rm p}^{\rm TH+FD}=0.74$ are calculated at $T=160$ MeV. At 200 MeV, the proportionality factor rises by 3\% to 2.95 and a decrease to 140 MeV reduces the density by $\sim18$\%. The density of observable protons is nominally a good proxy of the total thermal baryon density given that it amounts to approximately  1/3 of its value and varies slowly with the temperature of the system. However, decays are stochastic processes and the yield of their end product shall fluctuate. Let us for instance assume a system volume of 1000 fm$^3$ and a temperature of $T=160$ MeV. The thermal yield of protons should then be of order of $0.0036\times 1000=3.6$ per event while the $\rm TH+FD$ contributions amounts to average of $0.0114\times 1000=11.4$, an increase by a factor of three. The standard deviations of the FD contributions, $\sim \sqrt{7.8} = 2.8$ to the total proton yield are then almost as large as the thermal yield itself. This evidently neglects losses at low transverse momenta expected in practical measurements as well as effects of fluctuations of the number of thermal protons and correlations with anti-baryons due to production processes (baryon number conservation).  The proton yield is thus not a particularly precise indicator of the actual baryon number. 

It is also worth considering that the impact of FDs is expected to decrease as the temperature diminishes. At large collision energy, the initial temperature of large collision systems has been estimated based on real and virtual photons and exceeds 300 MeV~\cite{Shen:2013vja,STAR:2024bpc,ALICE:2015xmh}. At top RHIC and LHC energies, a large number of particles are produced in A--A collisions, and systems are estimated to live as long as 10 fm/$c$.  They eventually kinetically freeze out at temperatures below the pseudo-critical temperature of the phase transition. At small collision energy (RHIC BES), it is conceivable that the system  temperatures may in fact remain near or below the chemical freeze temperature, therefore greatly impacting the contributions from hadron decays. A thermal hadron gas at low temperature would indeed feature much smaller FD contributions. Net proton fluctuation cumulants might then experience variations that are driven by a reduction of FD decays rather than a phase transition. 

The computation of FD to (correlated) pair of stable particles proceeds similarly as the single particle case. One must, however, account for the fact that two stable species $a$ and $b$ can be produced as ``primary daughters", i.e., directly from a parent $\alpha$, or as ``cousins" from the sequence of decays ending into stable particles. The probability of decay into two stable species $a$ and $b$ is then computed iteratively starting with low $\alpha$ according to 
\begin{align} 
\label{eq:pairProb}
    P^{\alpha}_{ab} &= \sum_{i=1}^{n} B_{i}^{\alpha} P^{i}_{ab} + \sum_{i,j=1}^{n} B_{ij}^{\alpha} P^{i}_{a}P^{j}_{b}, 
\end{align}
where $P^{i}_{ab}$ is the probability of species $i$ to decay jointly (directly) into species $a$ and $b$ and $B_{ij}^{\alpha}$ is the branching fraction of species $\alpha$ into a  decay channel that involves species $a$ and $b$. Feed-down pair densities are computed according to 
\begin{align} 
\label{eq:pairFDDensity}
    \rho^{\rm FD}_{ab} &= \sum_{\alpha} \rho^{\rm TH}_{\alpha} P_{ab}^{\alpha}. 
\end{align}
Pair densities of species $a$ and $b$ are then the sum of uncorrelated thermal single density products and contributions from FD:  
\begin{align} 
\label{eq:pairTHFDDensity}
    \rho^{\rm TH+FD}_{a,b} &= \rho^{\rm TH}_{a}\rho^{\rm TH}_{b} + \rho^{\rm TH}_{a}\rho^{\rm FD}_{b} \\ \nonumber &+ \rho^{\rm FD}_{a}\rho^{\rm TH}_{b} + \rho^{\rm FD}_{ab}. 
\end{align}

\begin{figure}[hb]
\centering
\includegraphics[width=0.99\linewidth,trim={14mm 15mm 5mm 10mm}]{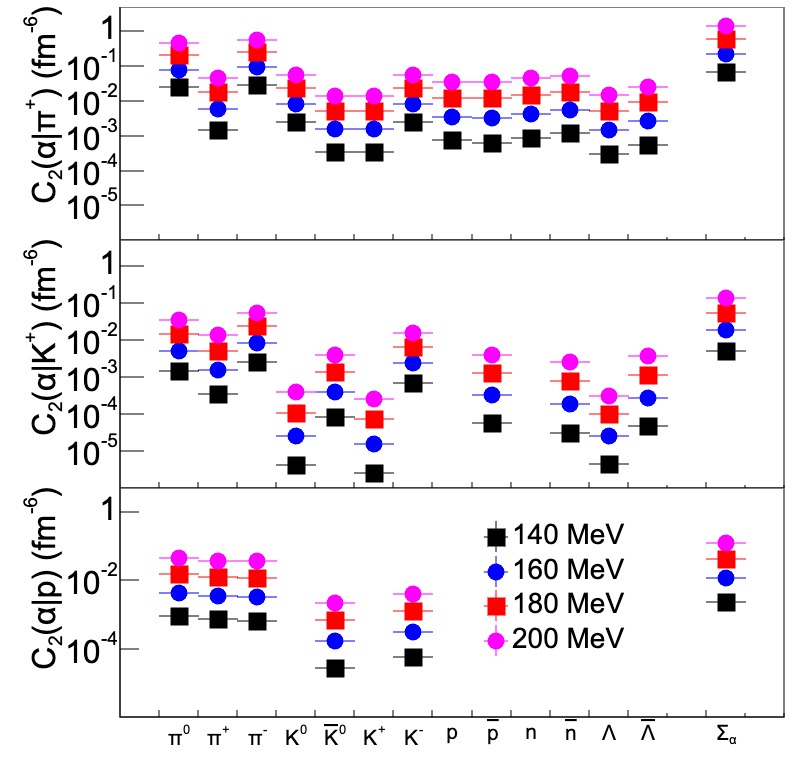}
\caption{Correlated pair densities of stable hadrons $\alpha$ emitted jointly with pions (top), kaons (middle), and protons (bottom) at selected temperatures. The right-most entries, labeled $\Sigma_{\alpha}$ correspond to the sum $\sum_{\alpha} C_2(\alpha|\beta)$ for $\beta = \pi^+, K^+, p$. }
\label{fig:StableHadronPairDensities}
\end{figure}

Given our interest in correlated pair decay FD, we  focus our discussion on two-particle cumulants defined as $C_2^{FD} = \rho_{ab} - \rho_{a}\rho_{b}$ that reduce to 
\begin{align} 
\label{eq:pairFDCumulant}
    C_2^{\rm FD}(a|b) &=  \rho^{\rm FD}_{ab} = \sum_{\alpha=1}^{n} \rho^{\rm TH}_{\alpha} P_{ab}^{\alpha}. 
\end{align}
The cumulants $C_2^{FD}(a|b)$ are straightforwardly computed with Eq.~(\ref{eq:pairFDCumulant}) but we further narrow our discussion to densities of pairs associated with a positive pion $\pi^+$, a positive kaon $\rm K^+$, or a proton $\rm p$ shown in the top, middle and bottom panels of Fig.~\ref{fig:StableHadronPairDensities}, respectively.

\begin{figure}[hb]
\centering
\includegraphics[width=0.99\linewidth,trim={6mm 20mm 10mm 10mm},clip]{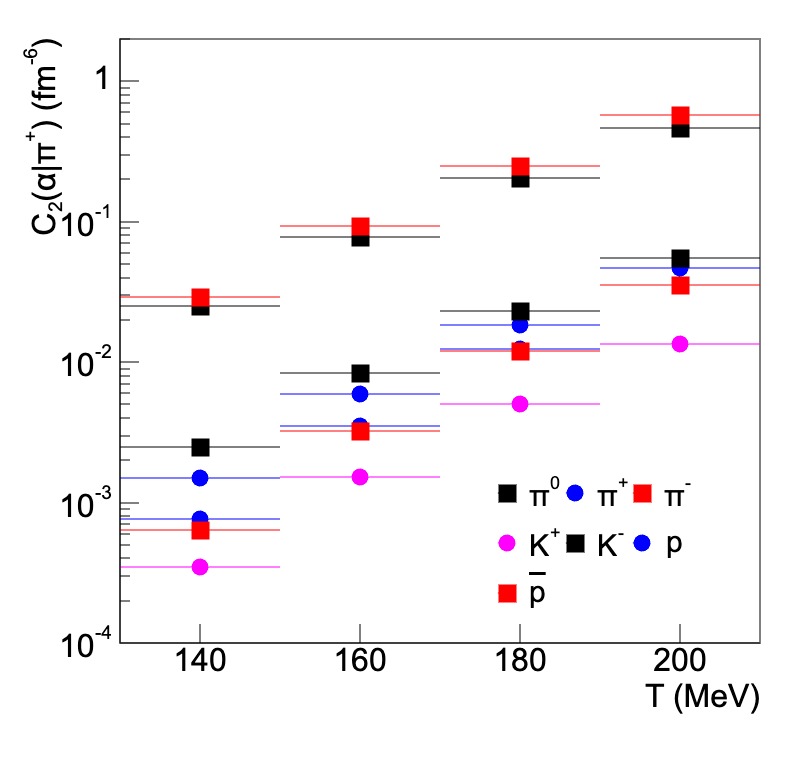}
\caption{Evolution with the hadron gas temperature of the correlated pair densities of stable hadrons $\alpha$ emitted jointly with a $\pi^{+}$. }
\label{fig:StableHadronPairDensitiesVsTemp}
\end{figure}

The emission of a $\pi^+$ is most often accompanied by (i.e., correlated to) a neutral ($\pi^0$) or a negative ($\pi^-$) pion, but correlated emission with a second $\pi^+$ is relatively strong compared, for instances, with the correlated emission of kaons and baryons. 
Correlated $\pi^+\pi^+$ emissions arise from the decay of excited (positive) states, e.g., $\pi^{+}(1800)\rightarrow \eta^0(547) + \eta' + \pi^+$ followed by  $\eta'\rightarrow \eta + \pi^+ + \pi^-$. Similarly, decays of excited (negative) states can also yield correlated $\pi^-\pi^-$ emissions. Figure~\ref{fig:StableHadronPairDensitiesVsTemp} displays the evolution of the correlated pair densities $C_2(\beta|\pi^+)$, due to decay FD, as a function of the temperature and clearly shows that contributions from decays increase rather significantly with the temperature. Given high mass hadrons have lifespan of the order 1 fm/$c$, similar to the lifespan of the hadron gas phase, it is unclear whether an expanding hadron gas can maintain thermal equilibrium as it rapidly cools down. As such higher mass states might not remain in equilibrium as long as lower mass states. Their ``earlier" decays would modify the trend displayed in Fig.~\ref{fig:StableHadronPairDensitiesVsTemp}. However, we have not attempted to model such a dependence. 

A range of mixed correlations also occurs with the production of a positively charged kaon ($\rm K^+$), as illustrated in the middle panel of Fig.~\ref{fig:StableHadronPairDensities}, which shows the amplitude of cumulants $C_2(\alpha|\rm K^+)$. One notes in particular that while the production of a $\rm K^+$ has a strong probability of being balanced by the emission of $\rm K^-$, the emission of a correlated $\rm K^+$ remains finite although weaker by more than two orders of magnitude. The emission of correlated $\rm K^+ K^-$ pairs is evidently primarily driven by the decay of the $\phi(1020)$ meson but other higher mass states also contribute, including for instance $\phi(1680), \phi_{3}(1850)\rightarrow \rm K^++K^-$, $f_2(2010), f_2(2340)\rightarrow \phi(1020) + \phi(1020)$, etc. It is additionally worth noticing that $\rm K^+$ are more likely to be charge balanced by the emission of one $\pi^-$ although decay induced correlations with $\pi^0$ and $\pi^+$ are also rather strong, whereas correlations with baryons are relatively weaker. 

Correlations involving one proton are arguably the simplest. No resonant state decay prominently into baryon anti-baryon pairs. However,  the annihilation of $\rm c\bar{c}$ states, such as $\eta_{\rm c}$, $\rm J/\psi$, etc, do decay into baryon anti-baryon channels plus other particles, albeit with  branching fractions of order $10^{-3}$ or smaller. The thermal production of $\rm c\bar{c}$ states is at best extremely weak given their large mass. Their decays thus contribute negligibly to the baryon yield and fluctuations of the net baryon number on an event-by-event basis. Consequently, as shown in the bottom panel of Fig.~\ref{fig:StableHadronPairDensities}, proton from decays are  most often associated  to pions, with the strength of correlations being the largest for neutral pions $\pi^0$ followed closely by that of $\pi^+$ and $\pi^-$.  Again, one notes that while (E\&M, weak and strong) decays do conserve baryon numbers, they do not necessarily yield charge and baryon balancing pairs.  We also observe that similar effects play out for correlations involving $\Lambda$ baryons.

Based on Fig.~\ref{fig:StableHadronPairDensities}, one finds that the correlated pair densities of many  species far exceed the expectation value of uncorrelated pair yield. For instance, the $T=160$ MeV thermal yield of $\pi^+$ and $\rm p$ are respectively $\rho_1^{\rm TH}(\pi^+)=0.0426\, {\rm fm}^{-3}$ and $\rho_1^{\rm TH}(\rm p)=3.66\times 10^{-3}\, {\rm fm}^{-3}$, thereby yielding a pair thermal density $\rho_1^{\rm TH}(\pi^+)\rho_1^{\rm TH}(\rm p)=1.56\times 10^{-4}\, {\rm fm}^{-3}$ which is a factor of $\sim 23$ smaller than the correlated two-cumulant $C_2(\rm p|\pi^+)=3.7\times 10^{-3}\, {\rm fm}^{-3}$. Likewise the thermal pair yield of $\rm K^+$ and $\rm K^-$, at the same temperature, amounts to $\rho_1^{\rm TH}(\rm K^+)\rho_1^{\rm TH}(\rm K^-)=1.88\times 10^{-4}\, {\rm fm}^{-3}$, whereas $C_2(K^-|K^+)=4.02\times 10^{-4}\, {\rm fm}^{-3}$, which is larger by a factor of $\sim 12$. One then concludes that even at the modest temperature of $T=160$ MeV, the pair correlated yields associated with decays typically dominate over thermal yields. Additionally note that the predicted correlated densities of $\rm p,\pi^+$ and  $\rm n,\pi^+$, as well as $\bar{\rm p},\pi^+$ and  $\bar{\rm n},\pi^+$ have approximately equal values. So while the decay of baryon does conserve the baryon number, it does not guarantee the daughter baryon has the same charge as the parent baryon. The moments  of net baryon number fluctuations are thus indeed potentially greatly  impacted by decays. The net proton number fluctuations cumulants  are  thus in fact a poor proxy of their full baryon number counterpart. 

\begin{figure}[ht]
\centering
\includegraphics[width=0.99\linewidth,trim={6mm 2mm 1mm 1mm},clip]{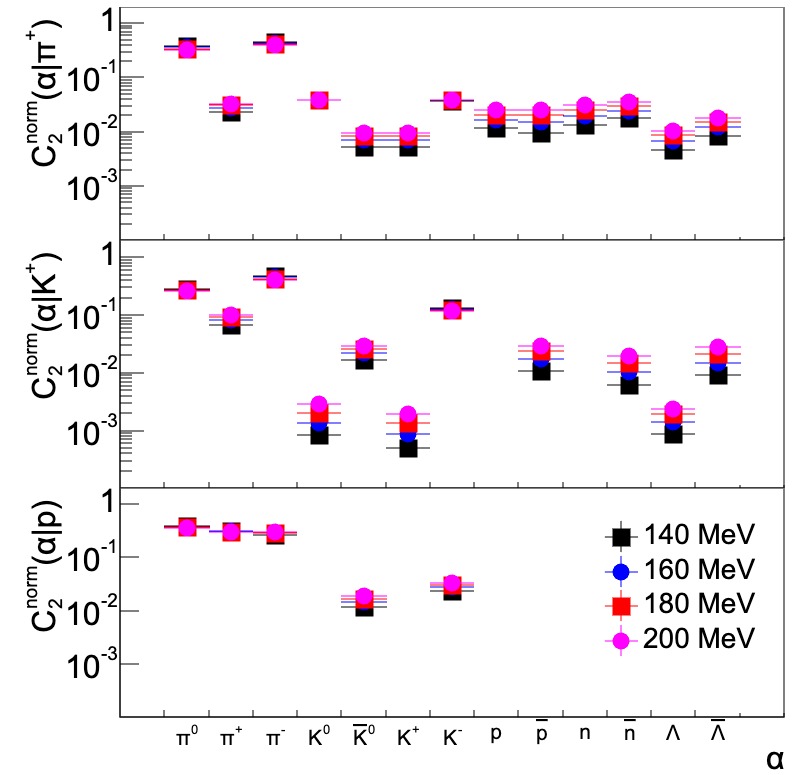}
\caption{Normalized correlated pair densities of stable hadrons $\alpha$ emitted jointly with pions (top), kaons (middle), and protons (bottom) normalized by their respective  integrals.}
\label{fig:StableHadronNormPairDensities}
\end{figure}

Within the context of processes driven by the strong interaction, conservation laws dictate that the baryon number, the charge and more generally the flavor content of produced hadrons must be balanced. However, charge, flavor, and baryon balance are not required to be realized by the correlated emission of only two particles. It is then possible for the emission of a proton to be charge balanced by the production of $\pi^-$ even though this does not satisfy the baryon balance requirement. It is  indeed possible to balance the charge, the baryon number, and quark flavor independently. This leads to simple sum rules of the form $\sum_{\alpha} B(\alpha| \beta)=1$, where $B(\alpha| \beta)$ is the integrated balance function of species $\beta$ being balanced by the emission of species $\alpha$~\cite{Pruneau:2022mui,Pruneau:2024jpa}. The production of a hadron $\beta$ can be charge balanced (or alternatively baryon balanced, etc) by several distinct hadrons and the sum of the integral of the balance functions must add to unity.  Such simple sum rules are evidently violated in the context of the GCE given (a) the emission of thermal hadrons are uncorrelated and (b) decays induces correlations that do not balance charge, baryon number, or quark flavors (although they do individually conserve quantum numbers). It is thus of interest to explicitly examine the impact of decays on balance functions by normalizing the 2-cumulants $C_2(\alpha|\beta)$ by their sum $\sum_{\alpha}C_2(\alpha|\beta)$, as illustrated in Fig.~\ref{fig:StableHadronNormPairDensities}.

Given the thermal particle production does not constrain the baryon or charge of produced particles, the probability of emitting charged baryons of the same mass but different charges are equal (with $\mu_Q=0$), the thermal yield of resonances such as $\Delta^{++}$, $\Delta^{+}$, $\Delta^{0}$, and $\Delta^{-}$ and their excited states are thus equal. As they decay, these resonances yield either a proton or a neutron in conjunction with one or two pions and thus indeed introduce fluctuations in the yield of protons vs. neutrons. This is for instance the case of the $\Delta^+(1232)$-baryon, which can decay either into ${\rm n}+\pi^+$ or ${\rm p}+\pi^0$, or the  $\Delta^0(1232)$-baryon, which may yield ${\rm p}+\pi^-$ or ${\rm n}+\pi^0$ decays. Heavier $\Delta$ and $\rm N^*$ states have similar competing decays into protons and neutrons and add to $\rm p$ vs $\rm n$ yield fluctuations. 

The thermal model does not ab initio feature correlations. One thus cannot here formulate a precise statement about the impact of correlations associated to decays in heavy collisions where medium to long range (rapidity/momentum) correlations associated with particle production dynamics are expected. But it is nonetheless clear that the magnitude of correlations associated with decays can be very large and might significantly impact both fluctuations of net quantum numbers as well as the amplitude and shape of balance functions. 

In the context of measurements of charge balance functions, one anticipates that the production of a $\pi^+$ ($\pi^-$) should be accompanied by a negative (positive)  hadron, and most often a $\pi^-$  ($\pi^+$). This basic expectation is however strictly violated if a sizable production (thermal or not) of heavy hadrons whose decay chains yield two $\pi^+$s or other species pair combinations. To illustrate this point quantitatively, Fig.~\ref{fig:StableHadronNormPairDensities} (top) presents the pair density associated to $\pi^+$ normalized by the total yield of pairs involving at least one $\pi^+$. One notes indeed that $\pi^+$s are   associated with highest probability with a $\pi^-$ but $\pi^0$ are close second, while the yield of associated $\pi^+$, i.e., a second $\pi^+$, remains at the level of a few percent. Decays also yield correlations between $\pi^+$ and kaons, as well as baryons and anti-baryons. The computation presented in this work assumes thermal populations and perfect/full measurement acceptance. They are thus not a precise predictor of the impact of the decays in A--A measurements at RHIC or LHC which must contend with finite transverse momentum and pseudorapidity acceptance, and as such need to extrapolate measured yields to vanishing $p_{\rm T}$. A more precise statement about the quantitative impact of FDs will indeed require a more detail model which explicitly generate $p_{\rm T}$ spectra and accounts for experimental limitations.  

\begin{figure}[ht]
\centering
\includegraphics[width=0.99\linewidth,trim={2mm 15mm 5mm 2mm},clip]{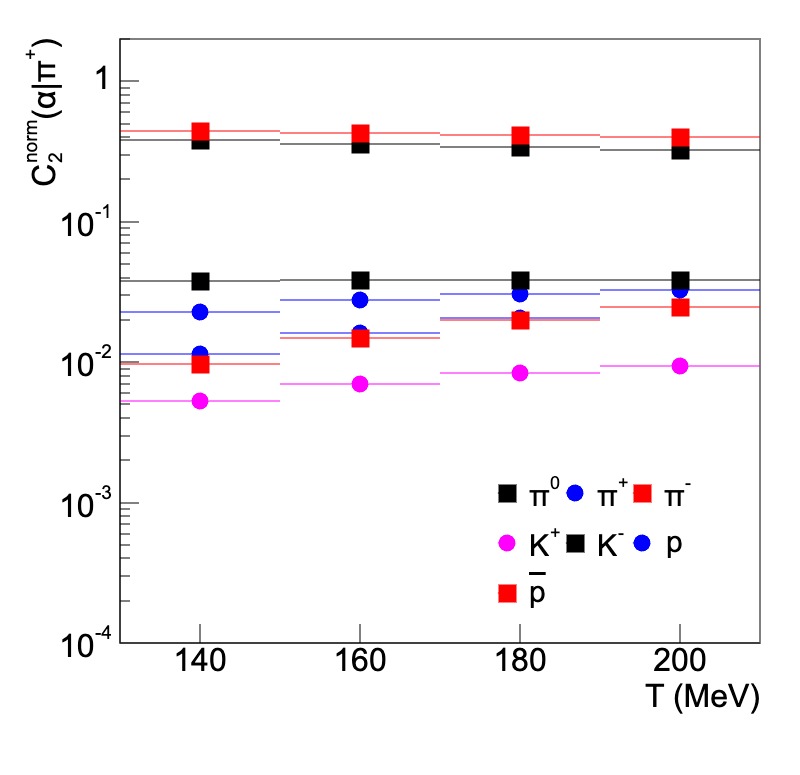}
\caption{Evolution with the hadron gas temperature of the normalized correlated pair densities of stable hadrons $\beta$ emitted jointly with a $\pi^{+}$. }
\label{fig:NormStableHadronPairDensitiesVsTemp}
\end{figure}

Figure~\ref{fig:NormStableHadronPairDensitiesVsTemp} displays a graph of the evolution with temperature of the normalized correlated pair densities, respectively, of stable hadrons. One observes, in particular,  that the normalized correlated densities $\pi^0+\pi^+$, $\pi^-+\pi^+$, exhibit small decreases, by 14.7\% and 9.7\%, respectively, while the normalized correlated densities of $\pi^++\pi^+$, ${\rm K}^++\pi^+$, ${\rm p}+\pi^+$ rises by 43\%, 77\% and 108\%, respectively, in that temperature range.  We conclude that fractional charge balancing pairs, e.g., $+-$, correlations are only weakly sensitive to the temperature, whereas non-charge balancing pairs from decays exhibit a stronger dependence of the chemical freeze-out temperature, owing to the increase of heavier resonances with rising temperature. These ``non-balancing" correlations thus potentially provide a better thermometer to assess the chemical freeze-out temperature of systems produced in A--A collisions. 

The analysis reported in this work is solely based on the thermal hadron gas model, which has had tremendous success in reproducing the relative abundances of individual yields of species produced in A--A collisions at both RHIC and LHC energies. One might however question the ability of the model to describe minimum bias or low multiplicity proton-proton collisions. It is thus interesting, before concluding this work, to examine relative contributions $\beta|\pi^+$  based on the \textsc{Pythia}8 Monte Carlo generator. It was found in Ref.~\cite{Pruneau:2024jpa} that \textsc{Pythia} 8.3 ``predicts" that in pp collisions at $\sqrt{s}=13$ TeV, a $\pi^+$ should be balanced (in full acceptance) by $\pi^-$, ${\rm K}^-$ and a $\bar{p}$ with respective fractions 88\%, $\sim 9$\% and $\sim 3$\%.
In Fig.~\ref{fig:StableHadronNormPairDensities}, all stable particles were included as potential correlated partners. Removing the contributions of neutral and positive particles, we find that fractions involving $\pi^-$, $K^-$ and $\rm \overline p$ accompanying a $\pi^+$ amount to  89\%, 8.0\% and 3.1\%, respectively, and thus  deviate only slightly from values obtained based on the \textsc{Pythia} model. One then concludes that the fractional integral of balance functions feature only modest sensitivity to the temperature of a thermalized system, and whether thermalization is in fact reached in these collisions.

\section{Conclusions}
\label{sec:conclusion}

In this work, we examined the impact of decays on the magnitude of fluctuations of net quantum numbers and the integral of balance functions based on a thermal hadron gas model. We found that the decay feed-down substantially impacts the yields of measurable ``stable particles" as well as correlated densities of species. Pion densities associated with decays far outshine those produced thermally and thus constitute a non-negligible source of net charge fluctuations. Additionally, although  the net  baryon number is preserved by the decays, the net proton number is not and can then fluctuate substantially because high mass baryons may yield neutron or lambda-baryon in addition to protons. Non-charge and non-baryon balancing effects associated with decays can also  be rather large. Finally, we also established that while charge balancing pair correlations do not feature a large sensitivity to the temperature, non-charge balancing do provide such a sensitivity given they are significantly impacted by decays. A brief comparison with predictions from \textsc{Pythia} 8.3 calculations additionally shows, however, that charge balance correlations (balance functions) are only somewhat sensitive to the degree of thermalization achieved in nucleus-nucleus collisions.

The results presented were based on a thermal model that features no built-in correlations and ideal observational conditions featuring to acceptance limitations, i.e., no limit on the rapidity acceptance and no minimum transverse momentum.  Further studies based on realistic models of the particle production including experimental constraints are thus needed to fully quantify the impact of decay FDs on measurements of net quantum number fluctuation cumulants and balance functions.

\section*{Acknowledgments}
The authors acknowledge the use of computer code and hadron data from the Therminator2 model and discussions with the creators of this model. The authors are also thankful for discussions with S. Voloshin and V. Vovchenko.  This work was supported in part by grant No. DE-FG02-92ER40713, No. DE-SC0021969, grant No. SR/MF/PS-02/2021-IITB  (E-37126), and grant No. PN-III-P4-PCE-2021-0390.
C.S. acknowledges a DOE Office of Science Early Career Award.

\newpage
\bibliography{reference.bib,ALICE.bib,BraunMunzinger.bib,ChunShen.bib,Pruneau.bib,Pratt.bib,STAR,Stephanov.bib}

% add this file for ALICE direct photons
%INSPIRE-Cite-1394677
\end{document}